\documentclass[prd,aps,showpacs,tightenlines,10pt, reprint,nofootinbib,superscriptaddress,floatfix,twocolumn, longbibliography]{revtex4-2}

\usepackage[utf8]{inputenc} 
\usepackage[T1]{fontenc}

\usepackage{acronym}
\usepackage{amsmath}
\usepackage{amsthm}
\usepackage{amssymb}
\usepackage{mathtools} 
\usepackage{phaistos}
\usepackage{bbm}

\usepackage{graphicx}
\usepackage{wrapfig}
\usepackage{subcaption}
\usepackage[section]{placeins}

\usepackage{physics}
\usepackage{IEEEtrantools}
\usepackage{xspace}
\usepackage{cancel}

\usepackage{epsfig}
\usepackage{multirow} 
\usepackage[dvipsnames,table,xcdraw]{xcolor}
\usepackage{rotating}
\usepackage[export]{adjustbox}
\usepackage{floatrow}
\usepackage{tabularx}
\usepackage{booktabs}
\usepackage{arydshln}
\usepackage{dcolumn}
\usepackage{bm}

\newfloatcommand{capbtabbox}{table}[][\FBwidth]

\usepackage[pdfencoding=auto, psdextra]{hyperref}
\usepackage{units}
\usepackage[normalem]{ulem}
\usepackage{orcidlink}
\usepackage{setspace}
\usepackage{verbatim}
\usepackage{float}
\usepackage{lipsum}

\graphicspath{{./figures/}}

\newcommand{\Msun}{\mathrm{M_\odot}}
\newcommand{\bvec}{\boldsymbol}

\allowdisplaybreaks

\begin{document}
\title{A New Probe of Dark Matter Subhalos: Stellar Aberration with TESS}
\author{Matthias Daniel\,\orcidlink{0009-0001-5805-2802}}
\thanks{Contact author: \href{mailto:daniel@itp.uni-frankfurt.de}{daniel@itp.uni-frankfurt.de}} 
\affiliation{Institute for Theoretical Physics, Goethe University, 60438 Frankfurt am Main, Germany}
\author{Xiao Xue\,\orcidlink{0000-0002-0740-1283}}
\thanks{Contact author: \href{mailto:xxue@ifae.es}{xxue@ifae.es}}
\affiliation{Institut de F\'{i}sica d’Altes Energies (IFAE), The Barcelona Institute of Science and Technology,
Campus UAB, 08193 Bellaterra (Barcelona), Spain}
\author{Kris Pardo\,\orcidlink{0000-0002-9910-6782}}
\affiliation{Department of Physics and Astronomy, University of Southern California, Los Angeles, CA 90089, USA}
\author{Laura Sagunski\,\orcidlink{0000-0002-3506-3306}}
\affiliation{Institute for Theoretical Physics, Goethe University, 60438 Frankfurt am Main, Germany}

\begin{abstract}
\noindent 
Small-scale dark matter (DM) structure encodes key information about the particle nature of DM and therefore provides a sensitive test of competing models. Yet, it remains hidden from electromagnetic surveys and is instead inferred through its gravitational effects. Stellar aberration, the apparent shift in a light source's position induced by the observer’s motion, offers a largely unexplored channel to access such signatures. DM subhalos can perturb the observer's motion, imprinting characteristic, spatially correlated shifts in stellar positions across the sky. We show that the Transiting Exoplanet Survey Satellite (TESS), with its long temporal baseline, wide sky coverage, and high-cadence observations, is well suited to search for these aberration signals. We derive Fisher-matrix-based sensitivity estimates for constant observer accelerations, forecasting a sensitivity down to $6.3\times 10^{-9}\,\mathrm{m/s^2}$ from the combined sample of TESS stars with magnitude $\mathrm{Tmag}\leq 10$. This sensitivity allows TESS to probe concentrated DM subhalos over a broad parameter space, from $\gtrsim 10^{-6}\,\Msun$ at AU-scale distances to $\gtrsim 10^{7}\,\Msun$ at $\mathcal O(10\,\mathrm{pc})$. TESS's sector-based observing strategy further provides intrinsic temporal resolution of potential DM-induced aberration signals. Moreover, we briefly discuss challenges for future data analysis, including the modeling of instrumental systematics and stellar astrometric foregrounds, such as parallax and proper motion. Our results establish stellar aberration as a novel probe of DM substructure, paving the way for dedicated searches in TESS and next-generation wide-field surveys.
\end{abstract}

\preprint{}
\maketitle

\textit{Introduction.}---Despite compelling evidence for dark matter (DM) from galaxy rotation curves \cite{Rubin_1978, Sofue_2001} and cosmological observations \cite{Springel_2006, Planck_2018}, its sub-galactic distribution remains largely unconstrained.

In the cold DM paradigm, hierarchical structure formation predicts subhalos down to Earth-mass scales \cite{Navarro_1996}. Their abundance, encoded in the subhalo mass function, depends sensitively on the underlying DM model \cite{Bringmann_2009}, with alternative scenarios such as warm, self-interacting, or ultralight DM generically suppressing small-scale structure or altering subhalo internal properties \cite{Bullock_2017, Bode_2001, Tulin_2018, Schutz_2020}. However, halos below a few $10^8\,\Msun$ are expected to lack luminous tracers due to inefficient star formation \cite{Kravtsov_2009, Efstathiou_1992, Hoeft_2006, Benitez_Llambay_2020}, rendering them invisible in electromagnetic surveys. Their detection therefore relies on gravitational probes, including lensing, pulsar timing, and stellar astrometry \cite{Buckley_2018, Bechtol_2023}.

Microlensing searches \cite{Rahvar_2015} target compact objects with masses $\sim10^{-10}-10^2\,\Msun$ \cite{Griest_2011, Li_2012, Niikura_2019_OGLE, Niikura_2019_Subaru, Verma_2023, Wyrzykowski_2023, Carr_2026}, while extended subhalos imprint lensing signatures sensitive to their internal structure \cite{Croon_2020, Croon_2020_HSC, Fedorova_2016}. Astrometric weak lensing reaches $\sim10^{-6}-10^9\,\Msun$ by measuring correlated positional shifts of background sources \cite{Tilburg_2018, Mondino_2020}, and Gaia analyses have constrained compact subhalos in the range $10^7 - 10^9\,\Msun$ \cite{Mondino_2020, Mondino_2024}. Future data releases and surveys, including Roman \cite{Roman_committee_2025}, will extend sensitivity toward lower masses \cite{Pardo_2021, Chen_2023, Fardeen_2024}. Pulsar timing arrays (PTAs) probe $\sim 10^{-13} - 10^3\,\Msun$ through timing residuals from Shapiro delays and Doppler shifts induced by accelerations of the Earth or pulsar \cite{Dror_2019, Seto_2007, Siegel_2007, Baghram_2011, Kashiyama_2018, Ramani_2020, Lee_2021_MOCK, Afzal_2023, Lee_2021}.

Stellar aberration is a well-established effect in astrometric measurements of stellar positions and proper motions \cite{Kovalevsky_2003, Liu_2013, Liu_2024, Brown_2025}. Unlike lensing or PTA signals, it produces global correlated shifts across the sky rather than localized distortions. A previous DM application considered observer oscillations induced by ultralight dark photon fields, with sensitivity to dark photon masses in the range $\sim 10^{-23} - 10^{-21}\,\mathrm{eV/c^2}$ \cite{Guo_2019}. Here, we instead study aberration induced by the observer's acceleration due to localized gravitationally bound structures, including both DM subhalos and baryonic matter distributions. Constraints on solar-system barycenter (SSB) accelerations have been obtained from Gaia astrometry and pulsar timing, reaching sensitivities at the $\sim10^{-10}\,\mathrm{m/s^2}$ level \cite{Gaia_EDR3_2021, Zakamska_2005, Titov_2013}. We demonstrate that the Transiting Exoplanet Survey Satellite (TESS) provides a powerful probe of DM subhalos over a broad parameter range by forecasting its sensitivity to constant accelerations via a Fisher-information approach.

\begin{figure*}[t!]
    \centering
    \includegraphics[width=\linewidth]{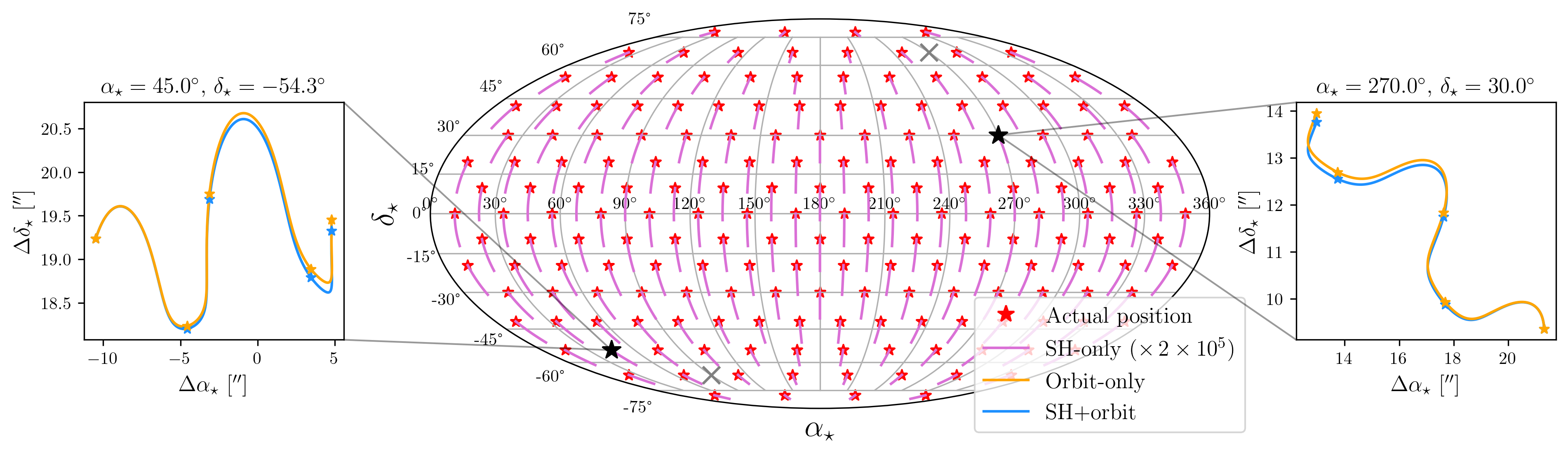}
    \caption{Stellar aberration during the first two operational TESS orbits (2018 July 25 -- August 22), computed from the spacecraft ephemerides \cite{jpl_horizons}. The Mollweide projection shows the SH-only dipolar aberration pattern for uniformly distributed star positions in right ascension ($\alpha_\star$) and declination ($\delta_\star$), rescaled by $2\times10^{5}$ ($M_\mathrm{SH}=5\times10^2\,\Msun$, $d_\mathrm{SH}=150\,\mathrm{AU}$). Gray crosses represent the ecliptic poles. Insets reveal stellar displacements, $(\Delta\alpha_\star,\Delta\delta_\star)$, with (blue) and without (orange) the additional SH-induced acceleration, where star symbols along the trail mark four equal time intervals and illustrate the rate of change of the aberration.}
    \label{fig:star_trails}
\end{figure*}

\textit{Theoretical background.}---Stellar aberration arises from the motion of an observer with velocity $\bvec{v}$ relative to incoming light from an astrophysical source. Let $\bvec{n}$ denote the unit vector pointing from the observer to the source in the observer's rest frame, such that the photon propagates along $-\bvec{n}$. For non-relativistic velocities, $|\bvec{v}|\ll c$, relevant to most solar-system dynamics, the apparent change $\delta \bvec{n}^\mathrm{ab}$ in propagation direction is given by the first-order approximation (FOA),
\begin{equation}
\delta \bvec{n}^\mathrm{ab} = \bvec{n}' - \bvec{n} = \frac{\bvec{v}}{c} - \left(\frac{\bvec{v}}{c}\cdot \bvec{n}\right)\bvec{n} = \frac{\bvec v_\perp}{c}\;,
\label{eq:first_order_aberration}
\end{equation}
where $\bvec{n}'$ denotes the Lorentz-boosted propagation direction and $c$ the speed of light in vacuum. The FOA depends only on the transverse velocity $\bvec{v}_\perp$, producing a deflection in the tangent plane of the celestial sphere with magnitude $\delta\theta^\mathrm{\,ab}\sim|\bvec{v}_\perp|/c$. For the Earth-Sun system, this corresponds to the well-known annual stellar aberration of $\sim20.5^{\prime\prime}$.

Since the observer is continuously moving, only temporal variations of the apparent position $\bvec{n}'$ are measurable. The aberration-induced contribution is
\begin{equation}
    \frac{d{\delta\bvec{n}^\mathrm{ab}}}{dt} = \underbrace{\frac{\bvec{a}}{c} - \left( \frac{\bvec{a}}{c} \cdot\bvec{n}\right)\bvec{n}}_{\text{change in velocity}} \, \underbrace{-\,\frac{1}{c}\left[\left(\bvec{v}\cdot \frac{d\bvec{n}}{dt}\right)\bvec{n} + \left(\bvec{v} \cdot \bvec{n}\right)\frac{d\bvec{n}}{dt}\right]}_{\text{change in line of sight}} \;,
    \label{eq:aberration_rate}
\end{equation}
where $\bvec{a} = d\bvec{v}/dt$ is the observer's acceleration at time $t$.

The second term accounts for changes in the aberration-free line of sight \cite{Liu_2013, Liu_2024, Brown_2025}, including parallax, stellar proper motion, spacecraft pointing jitter, and other instrumental or astrophysical systematics, discussed later. In the following, we focus on the first term, which captures changes in the observer’s velocity primarily induced by deterministic orbital motion as well as possible DM-driven perturbations.

Figure~\ref{fig:star_trails} illustrates this effect. We combine TESS's orbital motion, from the spacecraft ephemerides \cite{jpl_horizons, Giorgini_1996} for the first two operational orbits \cite{TESS_MAST} relative to the SSB in the International Celestial Reference Frame (ICRF), with the gravitational acceleration from a stationary SH with mass $M_\mathrm{SH}=5\times10^2\,\Msun$ placed at a distance $d_\mathrm{SH}=150\,\mathrm{AU}$ along the $-\hat{\bvec z}$ direction. These parameters are chosen for visibility and are not intended to be astrophysically realistic. The SH-only aberration signal (pink; amplified by $2\times10^{5}$) is dipolar across the sky, and the insets compare the apparent stellar displacements with (blue) and without (orange) the SH contribution. The SH produces a secular drift that accumulates on top of the aberration from TESS's orbital motion alone. Here, we show raw sky-plane trajectories; observational analyses require subtraction of common-mode field-of-view (FOV) motions, as discussed below.

\textit{TESS.}---TESS is a NASA mission launched in 2018 to conduct an almost all-sky photometric survey for transiting exoplanets \cite{Ricker_2014}. Four wide-field cameras with a combined FOV of $\sim 24^\circ \times 96^\circ$ survey the sky in 26 sectors of about $27.4\,\mathrm{days}$ each. The two-year primary mission achieved $\sim 85\%$ sky coverage, with up to $\sim 356.2$ days of continuous observations near the ecliptic poles due to sector overlap. TESS performs high-precision photometry for $>2\times10^5$ pre-selected bright stars at 2-minute cadence, complemented by full-frame images (FFIs) every 30 minutes, enabling observations of $\gtrsim 10^9$ targets in total. Pointing stability is maintained by an attitude control system supported by star trackers and science-image feedback, achieving sub-pixel accuracy with an angular resolution of $\sim 21^{\prime\prime}$ per pixel. The mission has completed its primary phase and is currently in extended operations, providing nearly eight years of data \cite{TESS_InstrumentHandbook, TESS_ObsGuide, Dichmann_2014}.

\begin{figure*}[t!]
    \centering
    \includegraphics[width=\linewidth]{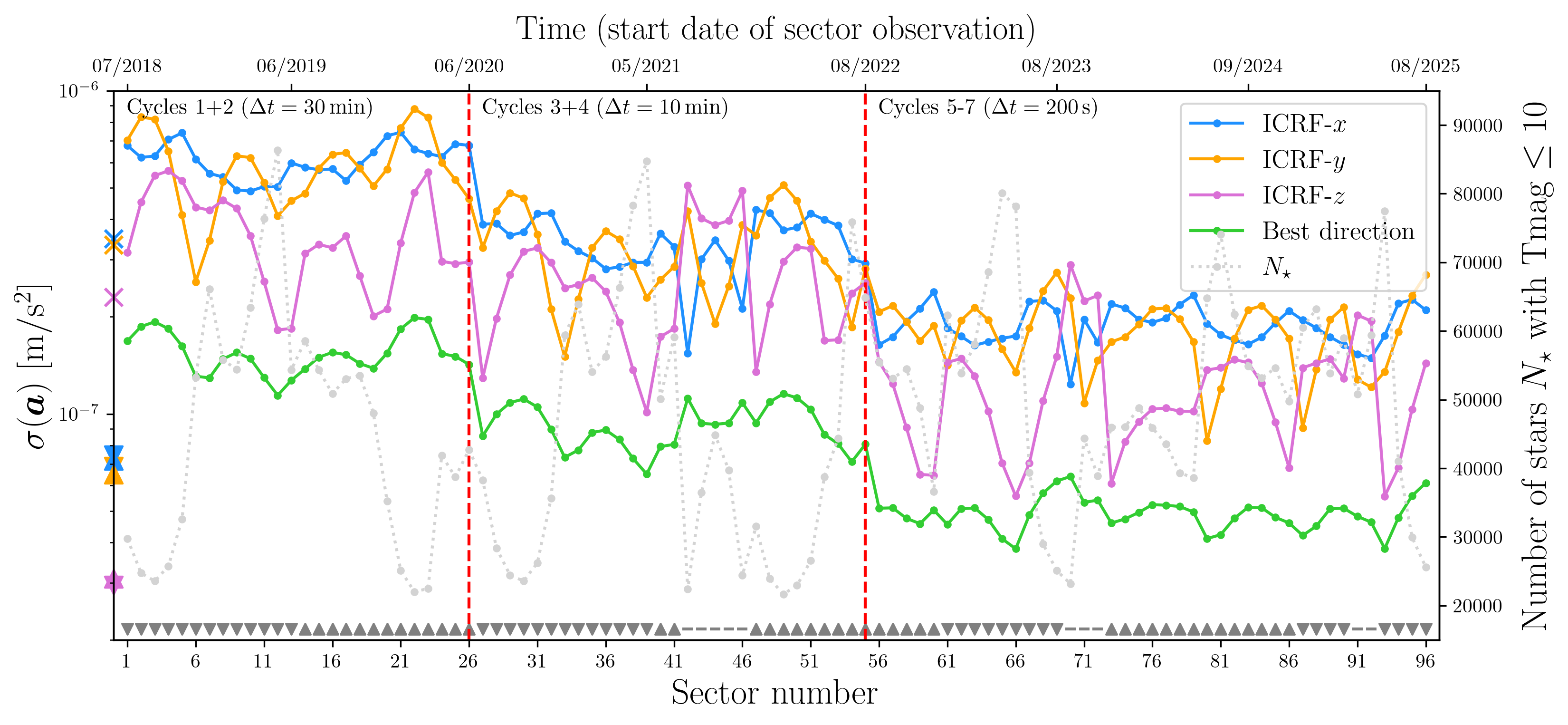}
    \caption{Fisher-matrix-based sensitivity estimates for a constant acceleration $\bvec a$ in the ICRF, using TIC stars with $\mathrm{Tmag} \leq 10$, including arbitrary FOV rotations and translations. Results are shown for the $x$-, $y$-, and $z$-components (blue, orange, and pink, respectively) across Sectors 1–96. Colored crosses on the left vertical axis indicate their mean values. The green curve presents the best sensitivity achievable for each sector. The upper horizontal axis shows the corresponding observing time span, highlighting the time-resolution capability of TESS. The dotted light gray line denotes the number of considered stars $N_\star$ per sector. Gray triangles at the bottom mark sectors contributing to the southern ($\textcolor{gray}{\blacktriangledown}$) and northern ($\textcolor{gray}{\blacktriangle}$) CVZs, used for the enhanced sensitivity estimates (colored triangles; Fig.~\ref{fig:NEP_SEP_region}). The remaining sectors correspond to ecliptic pointings and are marked by gray dashes. $\{T, \Delta t, \sigma\} = \{27.4\,\mathrm{days}, \{30\,\mathrm{min}, 10\,\mathrm{min}, 200\,\mathrm{s}\}, 0.21^{\prime\prime}\}$.}
    \label{fig:fisher_sensitivity_acc_rot+trans}
\end{figure*}

Although not designed as a precision astrometry mission, TESS's photometric centroid measurements can be exploited for astrometric analyses and searches for DM-induced stellar aberrations. Its large number of simultaneously observed bright stars provides strong statistical leverage while facilitating the identification of correlated spatial patterns and mitigation of common-mode instrumental effects. TESS provides large sky coverage per sector ($\sim 2,300\,\mathrm{deg}^2$), while Gaia achieves full-sky astrometry with a much smaller instantaneous FOV of $\sim 0.45\,\mathrm{deg}^2$ per telescope \cite{Gaia_2016}. This wide FOV enables a more uniform sampling of the dipolar aberration pattern and, together with repeated sector-based observations, offers long temporal baselines and multiple viewing geometries, particularly near the ecliptic poles. Overall, TESS combines high cadence, a large number of bright targets, wide sky coverage, and long-term orbital stability \cite{Gangestad_2013}, making it a complementary instrument for searching for DM-induced correlated centroid shifts.


\textit{Sensitivity estimates and parameter constraints.}---We forecast the per-sector acceleration sensitivity of TESS using a Fisher-information approach. We introduce an arbitrary acceleration vector $\bvec{a}$ in the ICRF, which causes a global, time-dependent aberration signal in stellar positions. This acceleration can be approximated as constant, as long as the considered observation time is much shorter than the timescale over which the subhalo's distance changes significantly. The signal corresponds to an additional perturbation on top of the deterministic TESS orbital motion \cite{jpl_horizons} [see Fig.~\ref{fig:star_trails}], without biasing the sensitivity estimate. To capture dominant instrumental systematics, we include arbitrary FOV rotations and translations in our analysis (see Appendix).

Assuming a uniform measurement uncertainty $\sigma$ across all stars and observation times, the relevant Fisher matrix $\hat{F}$ is
\begin{equation}
    \hat{F} = \frac{T^3}{12 \sigma^2 c^2\Delta t}\, \hat{X}\;,
    \label{eq:fisher_matrix}
\end{equation}
where $T$ denotes the observation time per sector, $\Delta t$ the FFI cadence, and $\hat{X}$ encodes the stars' distribution on the sky. The acceleration uncertainty $\sigma(a_i)$ is obtained from
\begin{equation}
    \sigma(a_i) = \sqrt{(\hat{F}^{-1})_{ii}}\;,
\end{equation}
showing that the sensitivity improves with longer observing baselines and denser sampling, scaling as $T^{-3/2}$ and $\Delta t^{1/2}$, respectively. 

For our fiducial estimate, we include all TESS Input Catalog (TIC) \cite{TIC_CTL} stars with magnitude $\mathrm{Tmag}\leq 10$ ($\sim 9\times10^5$ targets), adopt $\sigma = 0.21^{\prime\prime}$ ($10^{-2}\,\mathrm{pixels}$) as a conservative value for this magnitude range and motivated by empirical studies of TESS astrometric performance \cite{Gai_2022}, and take $T=27.4\,\mathrm{days}$ \cite{TESS_InstrumentHandbook}. We restrict our analysis to Sectors 1–96 (Cycles 1–7), excluding the incomplete Cycle 8 \cite{TESS_sectors}. The cadence improves from $\Delta t = 30\,\mathrm{min}$ during the primary mission to $200\,\mathrm{s}$ in later cycles \cite{TESS_InstrumentHandbook}. As shown in Fig.~\ref{fig:fisher_sensitivity_acc_rot+trans}, averaging over all considered sectors yields a sensitivity of $(3.5, 3.3, 2.3)^T \times 10^{-7}\,\mathrm{m/s^2}$ along the ICRF axes ($x$:~blue, $y$:~orange, $z$:~pink). The green curve indicates the best sensitivity achievable for each sector when allowing the acceleration direction to vary over the full sky. We note that the temporal structure of an aberration signal can be tracked across multiple sectors, enabling time-domain separation of local perturbers from stationary backgrounds.

The regions around the southern and northern ecliptic poles (SEP/NEP) are observed almost continuously. While the number of stars $N_\star$ in these fields is significantly lower than for a full sector (dotted light gray line), $\sigma(\bvec{a})$ benefits strongly from the extended observing baseline. Restricting the Fisher analysis to $\mathrm{Tmag} \leq 10$ stars near the poles shared by all $43$ sectors contributing to the continuous viewing zones (CVZs) per hemisphere, improves the sensitivity to $(7.5, 6.9, 3.0)^T \times 10^{-8}\,\mathrm{m/s^2}$ and $(7.2, 6.5, 3.1)^T \times 10^{-8}\,\mathrm{m/s^2}$ for the SEP (downward-pointing triangles) and NEP (upward-pointing triangles) region, respectively. However, the best sensitivities are reached in the directions $(92.0^\circ,-67.2^\circ)$ and $(272.2^\circ,64.9^\circ)$ for the SEP and NEP CVZs, respectively, yielding $\sigma(\bvec{a})=6.7\times10^{-9}\,\mathrm{m/s^2}$ and $6.3\times10^{-9}\,\mathrm{m/s^2}$ [see Fig.~\ref{fig:NEP_SEP_region}]. Extending the analysis to the full TIC sample ($\sim1.5\times10^9$ targets), together with future sector observations, offers further potential for improved sensitivity.

\begin{figure}[t]
    \centering
    \includegraphics[width=\columnwidth]{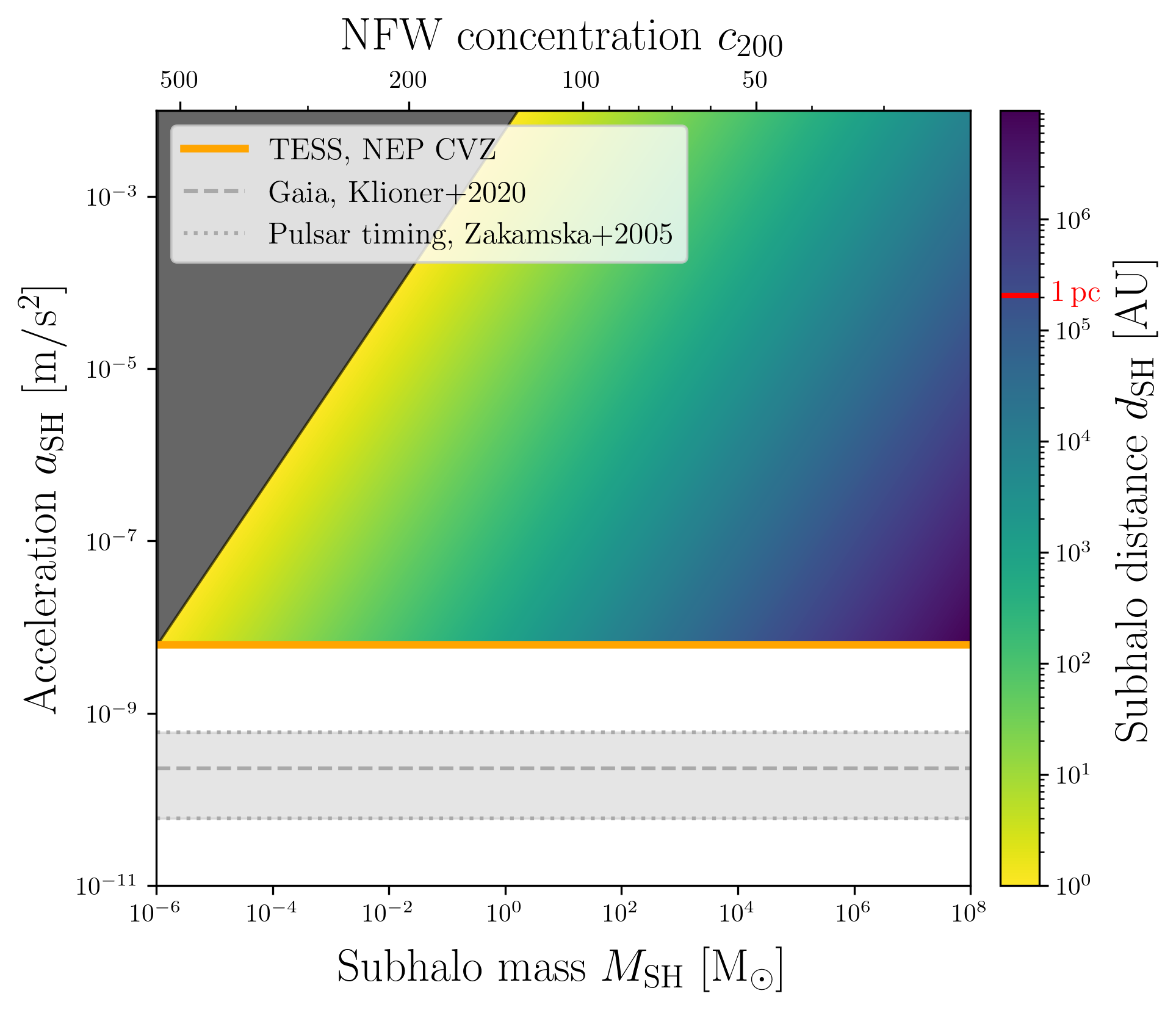}
    \caption{$(M_\mathrm{SH},a_\mathrm{SH})$ plane accessible to TESS based on the acceleration sensitivity for the NEP CVZ, $\sigma(\bvec a)=6.3\times10^{-9}\,\mathrm{m/s^2}$ (orange line). The color indicates the subhalo distance $d_\mathrm{SH}$, while the upper horizontal axis denotes the associated NFW concentration parameter $c_{200}$. For comparison, the best reported Gaia sensitivity \cite{Gaia_EDR3_2021} (dashed gray) and the range of pulsar timing sensitivities \cite{Zakamska_2005} (dotted gray) for SSB accelerations are shown, neglecting directional dependence. The black region corresponds to accelerations requiring distances below the lowest considered for a given mass.}
    \label{fig:subhalo_param_space_TESS}
\end{figure}

We translate this sensitivity into constraints on the DM subhalo parameter space by considering the acceleration $a_\mathrm{SH}$ caused by an individual subhalo. Requiring $a_\mathrm{SH}$ to exceed the most stringent Fisher threshold found, $\sigma(\bvec a)=6.3\times10^{-9}\,\mathrm{m/s^2}$ (NEP CVZ, orange line), determines the accessible region in the $(M_\mathrm{SH},a_\mathrm{SH})$ plane shown in Fig.~\ref{fig:subhalo_param_space_TESS}. The corresponding distance $d_\mathrm{SH}$ is represented by the color coding. This forecast reach extends from subhalos of $\gtrsim10^{-6}\,\Msun$ at AU scales to $\gtrsim10^7\,\Msun$ at $\mathcal{O}(10\,\mathrm{pc})$ distances. The black area designates accelerations that, at fixed mass, can only be induced by subhalos closer than the minimum considered distance. We also depict the Gaia \cite{Gaia_EDR3_2021} (dashed gray) and pulsar timing \cite{Zakamska_2005} (dotted gray) thresholds, using their published best-case sensitivities while neglecting directional dependence for an approximate comparison.

The upper axis in Fig.~\ref{fig:subhalo_param_space_TESS} indicates the inferred Navarro-Frenk-White (NFW) concentration parameter $c_{200}$ \cite{Navarro_1996} using the mass-concentration relation $c_{200}(M_{200}) = 10^{0.905} \left[M_{200}/(10^{12} h^{-1} \Msun)\right]^{-0.101}$ \cite{Dutton_2014}, where we set $M_{200} = M_\mathrm{SH}$ and $h = 0.674$ \cite{Planck_2018}. Although calibrated for field halos at redshift $z=0$ with $M_{200}\gtrsim10^9\,\Msun$, we extrapolate this relation heuristically into the low-mass regime \cite{Sanchez_Conde_2014}. The accessible parameter space corresponds to $c_{200}\approx 20-540$, consistent with expectations for dense substructure \cite{Sanchez_Conde_2014}, while even higher values ($\sim1.6\times10^3$) have been found from strong-lensing analyses \cite{Minor_2021}. Moreover, tidal stripping preferentially removes outer material, causing subhalos to become more concentrated than field halos of equal mass \cite{Diemand_2007}. Hence, the concentrations derived here should be regarded as conservative estimates.

Beyond individual subhalos, TESS's all-sky and sector-based strategy naturally enables population-level searches for DM substructure.

\textit{Challenges in data analysis.}---To isolate a potential DM-induced aberration signal and mitigate degeneracies with known astrometric effects, all contributions to the apparent displacement of a target must be carefully evaluated and consistently modeled. The observed sky position $\bvec n_I'$ of target $I$ can be written as
\begin{equation}
\begin{aligned}
    \bvec{n}'_{I}(t) &= \bvec{n}_{I,0} +
    \delta \bvec{n}_I^\mathrm{vel-ab}(t) + \delta \bvec{n}_I^\mathrm{los-ab}(t) + \delta \bvec{n}_I^\mathrm{pm}(t)\\[0.05cm] 
    &+ \delta \bvec{n}_I^\mathrm{plx}(t) + \delta \bvec{n}_I^\mathrm{sys}(t) + \delta \bvec{n}_I^\mathrm{astro}(t)\;,
\end{aligned}
\label{eq:all_contributions}
\end{equation}
where $\bvec{n}_{I,0}$ denotes the fiducial sky location and the remaining terms capture time-dependent deterministic and stochastic perturbations. The dominant deterministic contribution arises from TESS's orbital motion around the SSB, producing the standard velocity aberration $\delta \bvec{n}_I^\mathrm{vel-ab}(t)$ [cf.~Eq.~(\ref{eq:aberration_rate})] with amplitudes of $\mathcal{O}(10^{\prime\prime})$. Its evolution is approximately periodic but exhibits short-term deviations due to gravitational perturbations from the Earth-Moon-Sun system and more massive planets like Jupiter, which modulate the orbital elements. Accurate modeling therefore requires high-fidelity spacecraft ephemerides \cite{Folkner_2014, jpl_horizons} to maintain a well-controlled baseline signal. A DM-induced aberration signature may arise either from a resting DM subhalo [see Fig.~\ref{fig:star_trails}], continuously accelerating TESS relative to the SSB, or as a time-localized deflection produced by a passing subhalo. In both cases, the temporal behavior departs qualitatively from the orbital aberration pattern. The secondary aberrational term $\delta \bvec{n}_I^\mathrm{los-ab}(t)$ originates from line-of-sight variations. However, according to Eq.~(\ref{eq:aberration_rate}), it is suppressed by a factor $|\bvec{v}|/c \sim 10^{-4}$ relative to leading-order geometric effects, discussed next.

The star's transverse proper motion $\bvec{\mu}_I^\mathrm{pm}$ introduces a secular drift, $\delta \bvec{n}_I^\mathrm{pm}(t)=\bvec{\mu}_I^\mathrm{pm} t$, with typical amplitudes of $\sim 10^{-2}-10^{-1}\,\mathrm{^{\prime\prime}/yr}$ for nearby stars, reaching $\sim 10\,\mathrm{^{\prime\prime}/yr}$ for high-proper-motion sources, such as Barnard’s star \cite{GaiaDR3}. Parallax further produces a purely geometric, annually modulated displacement, $\delta \bvec{n}_I^\mathrm{plx}(t) \approx \left[\left(\bvec{r}(t) \cdot \bvec{n}_{I,0}\right) \bvec{n}_{I,0} - \bvec{r}(t) \right]/R_I$ valid for $|\bvec{r}| \ll R_I$, where $\bvec{r}$ denotes TESS's orbital position and $R_I$ the target's distance. For $R_I \approx 10 - 100\,\mathrm{pc}$, typical for TESS's exoplanet target sample \cite{NASA_TESS_Writers_Guide_2016}, this corresponds to $\sim {10^{-2}-10^{-1}}^{\prime\prime}$. Both effects are tightly constrained by Gaia astrometry \cite{GaiaDR3, Gaia_data, Lindegren_2021}, while extragalactic sources, such as quasars, exhibit negligible contributions.

The leading observational limitations arise from instrumental and astrophysical systematics. Instrumental effects, $\delta \bvec{n}_I^\mathrm{sys}(t)$, include, among others, spacecraft pointing jitter, optical distortions, scattered light, thermal fluctuations, point spread function variations, and saturation effects \cite{TESS_InstrumentHandbook}. Prior to correction, these can induce centroid excursions at the arcsecond level \cite{Gai_2022, TESS_InstrumentHandbook}. Standard TESS preprocessing mitigates many of these effects via pointing reconstruction, low-order detrending, and quality filtering of FFIs, particularly during momentum dumps and periods of elevated background. In particular, these procedures largely suppress common-mode or mean image motion while preserving spatial gradients across the FOV, which may contain DM-induced signatures. Additional pixel-level processing, including cosmic-ray mitigation, further reduces contamination. Residual systematics are expected to remain well below the pre-calibration level and can be incorporated statistically as an effective correlated noise contribution in future data analysis \cite{TESS_InstrumentHandbook}.

Astrophysical contributions, $\delta \bvec{n}_I^\mathrm{astro}(t)$, may result from unresolved binaries, stellar activity, blending with nearby sources, weak gravitational lensing, and related effects. While generally non-deterministic and time-variable, they typically differ from coherent, FOV-wide patterns like DM-induced aberration.


\textit{Conclusions.}---We introduce stellar aberration as a new avenue to search for DM subhalos. An acceleration of the observer induces a coherent dipolar pattern of apparent stellar displacements across the sky, providing a signature distinct from the localized distortions targeted by astrometric lensing and pulsar timing searches. Using TESS's observing characteristics \cite{TESS_InstrumentHandbook}, we forecast Fisher-matrix-based sensitivity estimates for constant observer accelerations, incorporating FOV rotations and translations as dominant global instrumental systematics. We assume uncorrelated Gaussian noise; red noise expected in observational data is not captured, potentially degrading sensitivity.

For TIC stars with $\mathrm{Tmag}\leq10$, TESS reaches its best sensitivity for the CVZs around the ecliptic poles, achieving $6.3\times10^{-9}\,\mathrm{m/s^2}$. In terms of individual perturbers, this sensitivity corresponds to DM subhalos with NFW concentrations $c_{200}\approx20-540$, spanning masses from $\gtrsim10^{-6}\,\Msun$ at AU-scale separations to $\gtrsim10^7\,\Msun$ at $\mathcal{O}(10\,\mathrm{pc})$. This acceleration limit comes within an order of magnitude of the effective SSB acceleration sensitivities from Gaia astrometry \cite{Gaia_EDR3_2021} and pulsar timing \cite{Zakamska_2005}. Importantly, however, TESS would enable sensitivity to transient and time-varying accelerations because of its high cadence. We leave further investigations of these signals to future work.   

Reaching the SSB acceleration sensitivity with TESS would require roughly two additional decades of $200\,\mathrm{s}$ full-frame observations of the CVZs. However, as discussed by Ref.~\cite{Gaia_EDR3_2021}, extragalactic quasars provide a cleaner reference frame than Galactic stars, whose intrinsic kinematics can obscure the signals of interest. The PLATO mission \cite{PLATO_Mission_Handbook} could substantially extend sensitivity: rescaling our Fisher estimate to its observing strategy ($2\,\mathrm{yr}$ of continuous pointing covering a $2,132\,\mathrm{deg^2}$ FOV with $\gtrsim 10^4$ targets providing $50\,\mathrm{s}$-cadence centroids), while assuming $0.015\mathrm{^{\prime\prime}}$ as single-measurement centroid precision, yields a forecast sensitivity down to a few $10^{-11}\,\mathrm{m/s^2}$ \cite{PLATO_Mission_Handbook, Nascimbeni_2025}. 

Stellar aberration offers a complementary probe to Doppler tracking, which is primarily sensitive to the line-of-sight component of the observer's velocity \cite{Asmar_2005, Zwick_2022}. Its coherent imprint across numerous stars provides multiple tracers of the same acceleration, thereby enhancing the detectability of weak signals.

More broadly, stellar aberration responds to any sufficiently nearby gravitating object that accelerates the observer, not exclusively DM subhalos. Candidate DM signals require consistency checks with complementary probes, while null results would constrain the abundance of local subhalos. Future analyses combining TESS centroid time series \cite{TESS_MAST} with precise spacecraft ephemerides \cite{jpl_horizons}, stellar proper motions and parallaxes \cite{GaiaDR3, Gaia_data}, and dedicated systematics modeling can exploit the characteristic dipolar aberration morphology to distinguish DM-induced accelerations from source-specific effects and instrumental noise. This establishes wide-field, high-cadence photometric surveys as precision accelerometers for local DM substructure.

\textit{Acknowledgements.}---We thank Yifan Chen for valuable discussions and constructive feedback on the manuscript. X.X.~is funded by grant CNS2023-143767, which is funded by MICIU/AEI/10.13039/501100011033 and the European Union NextGenerationEU/PRTR. This paper includes data collected with the TESS mission, obtained from the MAST data archive at the Space Telescope Science Institute (STScI). Funding for US Institutions for the TESS mission is provided by the NASA Explorer Program. STScI is operated by the Association of Universities for Research in Astronomy, Inc., under NASA contract NAS5-26555. This work makes use of the JPL Horizons on-line solar system data and ephemeris computation service. 

\textit{Data availability.}---The data and code supporting the findings of this letter are openly available \cite{Daniel_TESS}.

\bibliography{ref}

\clearpage
\onecolumngrid
\begin{center}
    {\large\bfseries End Matter\par}
    \vspace{0.15cm}
\end{center}
\twocolumngrid
\renewcommand{\thefigure}{EM\arabic{figure}}
\renewcommand{\theequation}{EM\arabic{equation}}
\setcounter{figure}{0}
\setcounter{equation}{0}

\begin{figure*}
    \centering
    \begin{subfigure}[t]{0.45\textwidth}
        \includegraphics[width=0.9\linewidth]{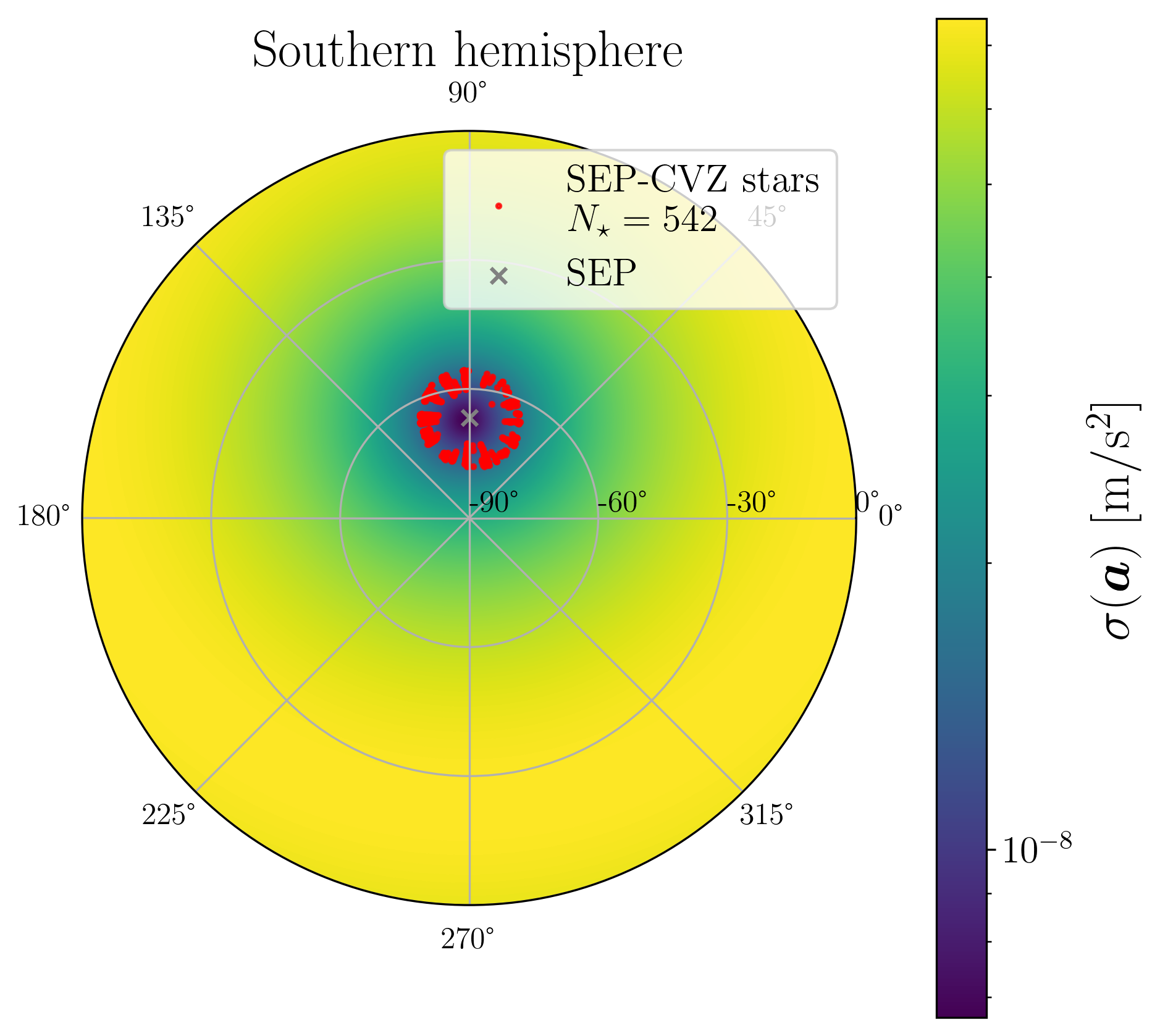}
    \end{subfigure}
    \hspace{0.3cm}
    \begin{subfigure}[t]{0.45\textwidth}
        \includegraphics[width=0.9\linewidth]{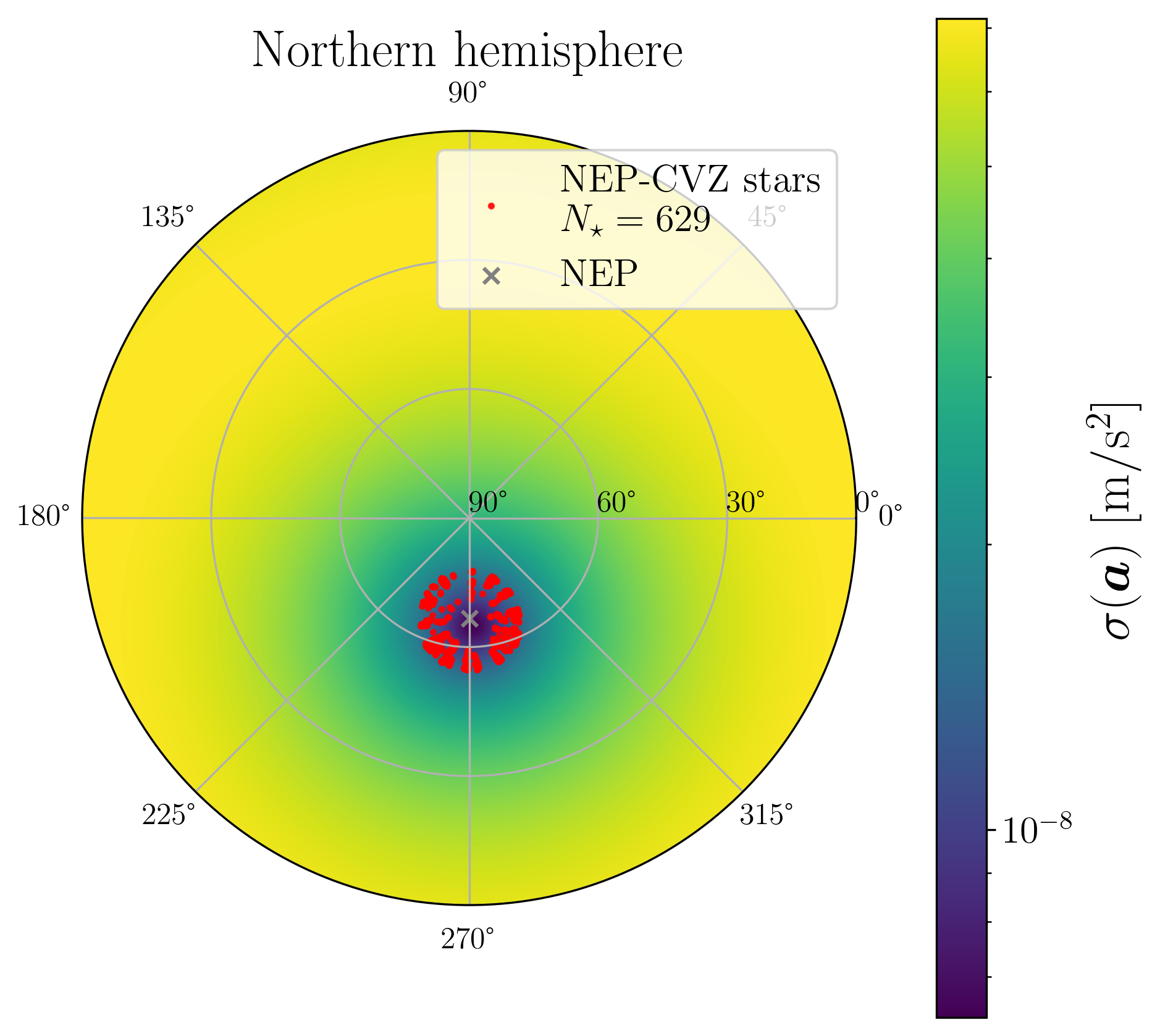}
    \end{subfigure}
    \caption{Color-coded acceleration sensitivity across the southern (left) and northern (right) hemisphere, derived from the $\mathrm{Tmag}\leq10$ stars (red dots) in the SEP and NEP CVZ, respectively. The ecliptic poles are marked by gray crosses. The Cartesian ICRF axes correspond to $x$: $\alpha = 0^\circ, \delta=0^\circ$; $y$: $\alpha = 90^\circ, \delta=0^\circ$; $z$: $\delta=90^\circ$. The sensitivity pattern is invariant under reversal of the acceleration direction.}
    \label{fig:NEP_SEP_region}
\end{figure*}
Here, we present a detailed derivation of the Fisher matrix from Eq.~(\ref{eq:fisher_matrix}). We compute the sensitivity $\sigma(\bvec a)$ of TESS to a constant acceleration $\bvec a$ in the ICRF basis $\{\bvec e_i\}$, starting from the global stellar aberration signal caused by the observer’s velocity $\bvec v(t)=\bvec v_0+\bvec a t$, with $\bvec v_0 = \mathrm{const}$. We further include FOV rotations and translations as the dominant global instrumental systematics in order to quantify their impact on the inferred acceleration sensitivity.\\[0.25cm]
\noindent \textbf{Physical model}\\[0.1cm]
In the FOA, the aberration-induced displacement of star $I = 1, \dots, N_\star$ at cadence $k = 1,\dots , K$ is given by
\begin{equation}
    \delta \bvec{n}^{\mathrm{ab}}_{I,k} = \frac{1}{c} \hat{P}_I \bvec{v}(t_k)\;,
\end{equation}
where $\hat{P}_I = \mathbbm{1}-\bvec{n}_{I,0}\bvec{n}_{I,0}^T$ projects onto the plane orthogonal to the fixed line of sight $\bvec{n}_{I,0}$ [see Eq.~(\ref{eq:first_order_aberration})]. In addition, $N_\star$ denotes the number of stars observed at each cadence and $K$ the number of individual observations, while $\mathbbm{1}$ is the $3\times 3$ identity matrix.

We model spacecraft translations by a constant vector $\bvec{\Delta}_k$ at each cadence, representing a FOV-mean subtraction that removes the average focal-plane displacement. This is standard in TESS centroid preprocessing and defines a star-independent shift,
\begin{equation}
    \delta \bvec{n}^{\mathrm{trans}}_{k} = \bvec \Delta_k\;.
\end{equation}

A cadence-dependent infinitesimal rotation around an arbitrary axis is parametrized by $\bvec{\omega}_k$, yielding
\begin{equation}
    \delta \bvec n_{I,k}^{\mathrm{rot}} = \bvec \omega_k \times \bvec n_{I,0} = - \hat{N}_{I,0}\, \bvec{\omega}_k\;,
\end{equation}
which is tangential to the celestial sphere and where
\begin{equation}
\hat N_{I,0} =
\begin{pmatrix}
0 & -n_z & n_y \\
n_z & 0 & -n_x \\
-n_y & n_x & 0
\end{pmatrix}_{I,0}\;.
\end{equation}\\[0.15cm]
\noindent \textbf{Measurement model}\\[0.1cm]
The observed stellar positions $\bvec{n}'_{I,k}$ at the $k$-th cadence are modeled as
\begin{equation}
\begin{aligned}
    \bvec{n}'_{I, k} = \bvec \mu_{I,k}(\bvec a, \bvec \Delta_k, \bvec \omega_k) + \bvec{n}_{I,0} + \frac{1}{c} \hat{P}_I \bvec{v}_0 + \bvec{\epsilon}_{I,k}\;,
    \label{eq:measurement_model}
\end{aligned}
\end{equation}
with
\begin{equation}
    \bvec \mu_{I,k}(\bvec a, \bvec \Delta_k, \bvec \omega_k) = \frac{t_k}{c} \hat{P}_I \bvec{a} + \bvec \Delta_k - \hat N_{I,0}\, \bvec \omega_k\;,
\end{equation}
and uncorrelated Gaussian noise $\bvec{\epsilon}_{I,k}$. We note that $\bvec n_{I,0}$ is known for each star, while the constant aberration term $\propto \bvec v_0$ is irrelevant for our analysis, as becomes clear below. Thus, the parameter vector is $\bvec \theta = (\bvec a, \{\bvec \Delta_k\}, \{\bvec \omega_k\})$ and the model is linear in all (3+3K+3K) parameters, which allows for an exact Fisher-matrix treatment.\\[0.25cm]
\noindent \textbf{Fisher matrix}\\[0.1cm]
Assuming a uniform uncertainty across all stars and observation times, i.e. $\sigma_{I,k} = \sigma$, the Fisher matrix reads
\begin{equation}
    \hat{F}_{\alpha \beta} = \frac{1}{\sigma^2} \sum_{I,k} \left(\frac{\partial \bvec{\mu}_{I,k}}{\partial \theta_\alpha}\right)^T \left(\frac{\partial \bvec{\mu}_{I,k}}{\partial \theta_\beta}\right)\;.
\end{equation}
The required derivatives are
\begin{equation}
\begin{aligned}
    \frac{\partial \bvec{\mu}_{I,k}}{\partial a_i} &= \frac{t_k}{c} \hat{P}_I \bvec{e}_i\;, \,\,\,\, \frac{\partial \bvec{\mu}_{I,k}}{\partial \Delta_{i,m}} = \delta_{km}\bvec{e}_i \;, \\[0.20cm]
    &\frac{\partial \bvec{\mu}_{I,k}}{\partial \omega_{i,m}} = - \delta_{km} \hat{N}_{I,0} \bvec{e}_i\;,
\end{aligned}
\end{equation}
where $\delta_{km}$ denotes the Kronecker delta.

Choosing the time origin such that the observation times are symmetric around the midpoint of the observing window, $\sum_k t_k = 0$, simplifies the Fisher analysis, as all cross-terms between the acceleration and cadence-independent parameters vanish. This justifies why the constant aberration term $\frac{1}{c}\hat P_I \bvec v_0$ in Eq.~(\ref{eq:measurement_model}) is not treated as an additional free parameter, even if it were in principle unknown. It remains uncoupled from the acceleration and therefore does not degrade the achievable acceleration sensitivity. This yields the following relevant block matrices:
\begin{equation}
    \begin{aligned}
        \hat{F}_{aa} &= \frac{1}{\sigma^2 c^2} \left(\sum_k t_k^2\right) \hat{M}\;, \,\,\,\, \hat F_{\Delta_k \Delta_m} = \frac{N_\star}{\sigma^2} \delta_{km} \mathbbm{1}\;, \\[0.15cm] 
        &\hat{F}_{\omega_k \omega_m} = \frac{\delta_{km}}{\sigma^2} \hat{M}\;, \,\,\,\, \hat F_{a \Delta_k} = \frac{t_k}{\sigma^2 c} \hat M\;, \\[0.15cm] 
        &\hat{F}_{a\omega_k} = -\frac{t_k}{\sigma^2 c} \hat{D}\;, \,\,\,\, \hat F_{\omega_k \Delta_m} = \frac{\delta_{km}}{\sigma^2} \hat N\;,
    \end{aligned}
\end{equation}
where $\hat{M} = \sum_I \hat{P}_I$, $N_\star = \sum_I 1$, $\hat{D} = \sum_I \hat{P}_I \hat{N}_{I,0}$, and $\hat N = \sum_I \hat N_{I,0}$.

To obtain the information on $\bvec a$, we marginalize over the nuisance parameters $\bvec n = (\{\bvec \Delta_k\}, \{\bvec \omega_k\})$. The marginalized Fisher matrix is given by the Schur complement,
\begin{equation}
\hat F_{aa}^{\mathrm{marg}} = \hat F_{aa} - \hat F_{an}(\hat F_{nn})^{-1}\hat F_{na}\;.
\label{eq:schur_complement}
\end{equation}

Since translations and rotations are independent for each cadence, the nuisance matrix $\hat F_{nn}$ is block diagonal, with a single-cadence nuisance block given by
\begin{equation}
\hat Q = \frac{1}{\sigma^2}
    \begin{pmatrix}
       \hat M & \hat N \\
        -\hat N & N_\star \mathbbm{1}
    \end{pmatrix}\;,
\end{equation}
such that $\hat F_{nn} = \mathrm{diag}(\hat Q,\dots, \hat Q)$. The coupling between acceleration and nuisance parameters at cadence $k$ reads
\begin{equation}
\hat F_{a n_k} = \frac{t_k}{\sigma^2 c}
\begin{pmatrix}
   -\hat D & \hat M
\end{pmatrix}\;.
\end{equation}
Thus, Eq.~(\ref{eq:schur_complement}) can be written as
\begin{equation}
\begin{aligned}
\hat F_{aa}^{\mathrm{marg}}& = \hat F_{aa} - \sum_k \hat F_{a n_k}\, \hat Q^{-1}\, \hat F_{n_k a}\\[0.15cm]
&= \frac{1}{\sigma^2 c^2} \left(\sum_k t_k^2\right) \hat X\;,
\label{eq:schur_complement_2}
\end{aligned}
\end{equation}
with 
\begin{equation}
    \hat X = \hat M -
\begin{pmatrix}
-\hat D & \hat M
\end{pmatrix}
\begin{pmatrix}
\hat M & \hat N\\
-\hat N & N_\star \mathbbm{1}
\end{pmatrix}^{-1}
\begin{pmatrix}
-\hat D^{T}\\
\hat M
\end{pmatrix}\;.
\end{equation}

For our sector-level analysis, we simplify the temporal contribution to $\hat F_{aa}^{\mathrm{marg}}$ by considering the limit of a large number of observations, $K = T/\Delta t \gg 1$, with uniform cadence $\Delta t$ and observation time $T$. Since $K \sim \mathcal{O}(10^3-10^4)$ for TESS, this approximation is well justified. This leads to
\begin{align}
    \sum_{k=1}^K t_k^2 \approx \frac{K T^2}{12} = \frac{T^3}{12 \Delta t}\;.
\end{align}

Finally, inserting this into Eq.~(\ref{eq:schur_complement_2}) results in Eq.~(\ref{eq:fisher_matrix}).\\[0.25cm]
\noindent \textbf{Cramér-Rao bound}\\[0.1cm]
The component-wise acceleration sensitivity ($i = x,y,z$) can be expressed as
\begin{equation}
    \sigma(a_i) = \sqrt{\left[\left(\hat F_{aa}^{\mathrm{marg}}\right)^{-1}\right]_{ii}}\;.
    \label{eq:Cramér-Rao}
\end{equation}

Compared to an aberration-only model, marginalizing over FOV rotations and translations degrades the sensitivity by average factors of about $(7.6, 6.6, 3.4)$ across the analyzed sectors.

Equation (\ref{eq:Cramér-Rao}) can be generalized to an arbitrary sky direction $\bvec u = (\cos(\delta) \cos(\alpha), \cos(\delta) \sin(\alpha), \sin(\delta))^T$, where $\alpha$ and $\delta$ denote right ascension and declination, respectively. It follows
\begin{equation}
    \sigma(a_{\alpha, \delta}) = \sqrt{\boldsymbol u^T \left(\hat F^\mathrm{marg}_{aa}\right)^{-1} \boldsymbol u}\;.
\end{equation}
Figure~\ref{fig:NEP_SEP_region} shows the corresponding results for the CVZs around the ecliptic poles.

We obtain the magnitudes and coordinates of the considered TIC stars from Ref.~\cite{TIC_CTL} and assigned them to the corresponding TESS sectors using the Python tool \texttt{tess-point} \cite{tess-point}. Our code is publicly available on GitHub \cite{Daniel_TESS}.

\end{document}